\documentclass[12pt,preprint]{aastex}
\begin{document}

\parindent=1.0cm

\title
{The Central Regions of M31 in the $3 - 5\mu$m Wavelength Region \altaffilmark{1}}

\author{T. J. Davidge \altaffilmark{2}}

\affil{Herzberg Institute of Astrophysics,
\\National Research Council of Canada, 5071 West Saanich Road,
\\Victoria, B.C. Canada V9E 2E7\\ {\it email: tim.davidge@nrc.ca}}

\author{Joseph B. Jensen}

\affil{Gemini Observatory, 670 North A'ohoku Place, 
\\Hilo, HI 96720-2700\\ {\it email: jjensen@gemini.edu}}

\author{K. A. G. Olsen}

\affil{Cerro Tololo Inter-American Observatory,
\\National Optical Astronomy Observatory, Casilla 603,
\\La Serena, Chile\\ {\it email: kolsen@noao.edu}}

\altaffiltext{1}{Based on observations obtained at the
Gemini Observatory, which is operated by the Association of Universities
for Research in Astronomy, Inc., under a co-operative agreement with the
NSF on behalf of the Gemini partnership: the National Science Foundation
(United States), the Particle Physics and Astronomy Research Council
(United Kingdom), the National Research Council of Canada (Canada),
CONICYT (Chile), the Australian Research Council (Australia), CNPq (Brazil),
and CONICET (Argentina).}

\altaffiltext{2}{Visiting Astronomer, Canada-France-Hawaii
Telescope, which is operated by the National Research Council of Canada,
the Centre National de la Recherche Scientifique, and the University of
Hawaii.}

\begin{abstract}

	Images obtained with NIRI on the Gemini North telescope are used to 
investigate the photometric properties of the central regions of M31 in the $3 - 5\mu$m 
wavelength range. The light distribution in the central arcsecond differs from 
what is seen in the near-infrared in the sense that the difference in peak brightness 
between P1 and P2 is larger in $M'$ than in $K'$; no obvious signature of P3 is detected 
in $M'$. These results can be explained if there is a 
source of emission that contributes $\sim 20\%$ of the peak $M'$ light of 
P1, that has an effective temperature of no more than a few hundred K and 
is located between P1 and P2. Based on the red $K-M'$ 
color of this source, it is suggested that the emission originates in a 
circumstellar dust shell surrounding a single bright AGB star. 
Tests of this hypothesis are described. A bright source that 
is $\sim 8$ arcsec from the center of the galaxy is also detected in $M'$. 
This object has red colors, and an absolute brightness in $M'$ that is similar to the 
most highly evolved AGB stars in the solar neighborhood; hence, it is 
likely to be a very evolved AGB star embedded in a circumstellar envelope. The $K-$band 
brightness of this star is close to the peak expected for AGB evolution, and 
an age of only a few hundred million years is estimated, which is 
comparable to that of the P3 star cluster. Finally, using high angular resolution 
near-infrared adaptive optics images as a guide, a sample of unblended AGB stars 
outside of the central few arc seconds is defined in $L'$. The $(L', K-L')$ 
color-magnitude diagram of these sources shows a dominant AGB 
population with a peak $L'$ brightness and a range of $K-L'$ colors that are similar to 
those of the most luminous M giants in the Galactic bulge.

\end{abstract}

\keywords{galaxies: individual (M31) - galaxies: stellar content - galaxies: nuclei -
stars: AGB and post-AGB}

\section{INTRODUCTION}

	The majority of nearby late-type spiral galaxies harbour distinct nuclei that 
often show evidence of recent star formation (e.g. Carollo et al. 2001; Boker 
et al. 2002; Scarlata et al. 2004). Because they are viewed at comparatively high 
spatial resolutions, the spiral galaxies in the Local Group and the nearest external 
ensembles, such as the Sculptor, Centaurus, and M81 groups, play a key 
role in understanding the nature of these nuclei. As the nearest 
external spiral galaxy, M31 is of particular importance 
for characterizing galaxy nuclei, and understanding the Galactic 
Center in the broader context of other systems.

	The gross morphological properties of M31 are similar to those of the 
Milky-Way, and the two galaxies have similar total masses. Nevertheless, aside 
from these broad similarities there have been significant differences in their 
evolutionary histories. Some of these differences may have originated during 
very early epochs; for example, the integrated spectra of M31 globular clusters 
show enhanced CN absorption when compared with Milky-Way globular clusters, 
and this may be due to differences in the early 
enrichment of the protogalactic environment (e.g. Beasley et al. 2004). 

	Some of the differences in evolutionary histories may also be a consequence of 
processes that continued to recent epochs, and possibly to the present day. There are 
indications that M31 is experiencing on-going tidal interactions with its companions 
(Ferguson et al. 2002, Zucker et al. 2004), and this is expected to have an impact on the 
global stellar content of the galaxy. Indeed, M31 appears to contain a population of 
intermediate age globular clusters that has no known counterpart in the Galactic 
cluster system, and the chemical content and dynamical properties of these clusters suggest 
an origin that is distinct from that of older M31 clusters 
(Beasley et al. 2005). There is also a population of stars with 
a disk-like metallicity that is seen well off of the plane of the M31 disk that appears 
to have an age of 6 Gyr, and the tidal stirring of the M31 disk and companions are one 
possible explanation for such a population (Brown et al. 2003). While the Milky-Way has also 
interacted with its companions (Ibata, Gilmore, \& Irwin 1994; Martin et al. 2004), these 
encounters evidently have not produced a large, intermediate-age, 
metal-rich extraplanar population. The total star formation rate (SFR) of M31 
at the present day is also significantly lower than that of the Milky-Way 
(0.6 versus 3 to 5 M$_{\odot}$ year$^{-1}$; Prantzos \& Aubert 1995; Williams 2003), 
although M31 likely had a higher SFR roughly 1 Gyr in the past (Williams 2002). 
Renda et al. (2005) investigate the very different chemical evolution histories of the 
Milky-Way and M31. 

	Differences between M31 and the Milky-Way are also evident when comparing 
the innermost regions of these galaxies. The central black hole 
in M31 is an order of magnitude more massive than that 
in the Milky-Way (e.g. Ghez et al. 2005; Bender et al. 2005). 
The central regions of the two galaxies are also morphologically different, in that while 
M31 contains two sources in the inner arcsec (e.g. Lauer et al. 1993), 
the center of the Milky-Way would appear as a single source when viewed at 
the same distance (Davidge et al. 1997b). The brightest of the central 
sources in M31 at wavelengths longward of $0.5\mu$m, P1, is spatially extended 
and offset from the isophotal center of the galaxy. The second source, P2, 
coincides with the isophotal center of the galaxy, and is more compact than P1. While 
the spectral energy distribution (SED) of P2 is such that it dominates over P1 in 
the near-ultraviolet (e. g. Lauer et al. 1998), both sources have similar 
near-infrared colors (Davidge et al. 1997a; Corbin, O'Neil, \& Rieke 2001). The 
Mgb and Ca triplet indices of P1 and P2 differ only slightly from those of the 
surrounding bulge (Kormendy \& Bender 1998). P1 and P2 
thus are not a single object riven by a dust lane, and 
Tremaine (1995) suggests a dynamical origin for these structures.

	Bender et al. (2005) used HST STIS spectra to investigate the stellar content of 
the central regions of M31, and isolated an A-type spectrum in the center of P2, which 
they call P3. P1, P2, and P3 appear to be dynamically coupled, consisting of nested disks 
(Bender et al. 2005). If modelled as a simple stellar system, P3 is a $\sim 200$ Myr 
population with a mass of 5200 M$_{\odot}$. The discovery of P3 is consistent 
with the analysis of lower angular resolution integrated spectra, which 
predicted a centrally concentrated young or intermediate age population in M31 (e.g. 
Bica, Alloin, \& Schmidt 1990; Davidge 1997; Sil'chenko, Burenkov, \& Vlasyuk 1998). 
For comparison, the Sgr A complex and its environs are sites of recent ($t < 10$ Myr) star 
formation (e.g. Cotera et al. 1996; Davidge et al. 1997a; Blum et al. 2003). There is 
also evidence for previous episodes of star formation in and around Sgr A, indicating 
that the current activity is not a unique occurence (Blum et al. 2003).

	Could the differences in nuclear stellar contents be related to the masses 
of the central black holes? Heckman et al. (2004) characterized the 
accretion and SFRs around black holes in a large sample of Type 2 AGNs, and found (1) that 
black-hole growth in the local Universe is fastest among black holes that 
have masses $< 10^8$ M$_{\odot}$, and (2) that the ratio of SFR to 
growth rate decreases towards higher black hole masses. 
Differences in the young stellar populations near the centers of bulges that harbour 
very large black holes and those of more modest size, such as that in the Galactic 
Center, might then be expected as part of the overall downsizing trend in the Universe. 
This being said, the statistics of blue nuclei in nearby galaxies suggest that 
star formation likely occurs periodically near the centers of late-type spiral 
galaxies during intervals $\leq 1$ Gyr (e.g. Davidge \& Courteau 2002), and so the age 
difference between the most recent episodes of star formation in Sgr A and P3 may not 
be significant.

	Additional insights into the stellar content near the center of M31 
can be obtained from observations at wavelengths longward of $2.5\mu$m. This wavelength 
region probes light from sources with characteristic temperatures $\leq 10^3$ K, 
such as might originate in circumstellar environments. The sources expected to have the 
strongest circumstellar emission at these wavelengths are very massive and/or highly 
evolved stars. Objects of this nature are relatively rare, and at wavelengths 
longward of $2.5\mu$m the contrast between the objects that show hot circumstellar emission 
and the main body of stars is greatly increased. Images in the $3 - 5\mu$m 
wavelength interval might thus provide a means of resolving individual 
stars, albeit those that are highly evolved and hence most difficult to model, in 
regions that are very crowded at visible and near-infrared wavelengths. 

	In the current paper, $L'$ and $M'$ observations of the central regions of M31, 
obtained with the NIRI imager on Gemini North, are combined with high angular 
resolution $JHK$ images that were obtained with the CFHT adaptive optics (AO) system. 
These data are used (1) to investigate the SED of the central arcsecond of M31 in the 
$1 - 5\mu$m wavelength interval, and (2) to search for bright sources outside 
of the nuclear regions. The observations and the procedures used to reduce the data are 
discussed in \S 2, while the SED of the central arcsecond in the $1 - 5\mu$m 
wavelength interval is investigated in \S 3. The properties of the brightest 
circumnuclear sources are discussed in \S 4, and a summary and discussion 
of the results follows in \S 5.
  
\section{OBSERVATIONS \& REDUCTION}

\subsection{NIRI $M'$ and Short Exposure $L'$ Observations}

	The central regions of M31 were observed during the nights of September 
3, 4, and 6 2005 as part of the queue-scheduled observing program for NIRI on Gemini 
North. The program identification number is GN-2005B-Q-59. The instrument was configured 
with the f/32 camera; the image scale is thus 0.022 
arcsec pixel$^{-1}$ and the total imaged field is $22.5 \times 22.5$ arcsec$^2$. 
The galaxy nucleus was positioned near the center of the NIRI science field.

	The $M'$ data were recorded on all three nights, as a 
series of on-source/background field pairs. Sixty co-added 0.4 
second exposures were recorded of the central regions of M31, and these 
were followed immediately by a similar set of observations of a background region 
located roughly 1 arcminute from the nucleus. After this pair of observations 
was completed, the telescope pointing was returned to the center of M31 to sample the 
next step in the five point dither pattern. After all five dither positions had been 
observed the cycle was re-started at the first dither position; the central position 
of the background field was also offset from that used in the previous observing 
sequence to facilitate the suppression of bright sources in the background field. The
total on-source integration time obtained in this manner is 3500 seconds. A standard star 
close to M31 on the sky was observed either immediately before or after the observations 
on each night. The seeing measured from the standard star was 0.4 arcsec FWHM, 
and the standard deviation in the photometric zeropoints is $\pm 0.04$ magnitudes. 

	Observations of the central regions of M31 in $L'$ were recorded 
on the night of September 6, 2005. The total integration 
time for these data is much shorter than that 
of the $M'$ data, due to the lower sky background in $L'$. Sixty co-added 0.5 second 
exposures were recorded at each of 5 different dither positions. 
As with the $M'$ observations, a background field located roughly 1 arcmin away 
was also observed immediately after each dither position. The total on-source 
integration time is 150 seconds. Only the central $768 \times 768$ 
pixels of the NIRI detector were read out for these observations. 
These data will be referred to as the `short exposure' $L'$ observations. 
A standard star located close to M31 on the sky was observed immediately before the 
$L'$ data were recorded, and the image quality was measured to be 0.5 arcsec FWHM. 

	While deeper $L'$ images are available (\S 2.2), the short exposure $L'$ data are 
used in the study of the central arcsec of M31 in \S 3 because they were recorded during 
the same epoch as the $M'$ data. This is an important consideration as 
the brightest objects at these wavelengths tend to be photometric variables. 
Moreover, the $M'$ and short exposure $L'$ data are subject to the same acquisition 
and reduction procedures; hence, these data form a homogeneous 
dataset for probing the nuclear regions of M31. 

	To reduce the $M'$ and short-exposure $L'$ data, the background 
field observation in each data pair was subtracted from the corresponding observation 
of the center of M31. The resulting background-subtracted images were found to contain
a residual thermal pattern arising from thermal variations between the two positions. 
A calibration frame to remove this residual emission was constructed by 
median-combining the background-subtracted data obtained on each night without spatial 
registration to correct for the dither offsets. While the subtraction of this calibration 
frame removes the residual thermal signature, the low spatial frequency contributions 
from the M31 bulge outside of the nuclear regions are also suppressed. The pixels in 
the resulting data were block averaged in $5 \times 5$ pixel$^2$ groups 
($0.11 \times 0.11$ arcsec$^{2}$) to produce images with 
angular sampling that better matches the seeing, and 
these block-averaged images were spatially registered to correct for the dither 
offsets. A final image was then constructed by median-combining the registered images.

\subsection{NIRI Long Exposure $L'$ Observations}

	Deep $L'$ images of M31 were recorded with NIRI on the night of November 2 
2003, as part of a study of surface brightness fluctuations at infrared wavelengths in 
M31 and M32 (Jensen 2006, in preparation). The program identification number is 
GN-2003B-Q-43. The data were collected using a 512${\times}$512 subarray and the 
f/14 camera, so that each image covers 25.6 arcsec on a side with 0.05 arcsec 
pixel$^{-1}$ sampling. Two hundred exposures of 0.151 second duration were co-added 
at each dither position. A similar set of exposures was taken of an empty sky field
located 16.5 arcmin northwest of the nucleus immediately after each 
M31 dither position was observed. The total exposure time was 2748.2 sec, 
giving a $S/N \geq 5$ pixel$^{-1}$ even near the edges of the field of view. 
Standard stars were observed, and the resulting photometric zero point is consistent
with that measured in September 2005. 

	The sky frames bracketing each M31 observation were 
averaged, and the result was subtracted from the M31
observation. The sky-subtracted data were then divided by a flat 
field image, which was constructed by normalizing the average of all the sky images. 
This flat fielding step was judged to be necessary because the longer exposure 
time of these data lowered the background noise level to the point where it was 
comparable to flat field variations; this is not the case in the $M'$ and 
short exposure $L'$ data, where background noise dominates. The flat 
fielded data were then spatially aligned and averaged to construct the final
image. The image quality in the combined frame is 0.35 arcsec FWHM.
The final image, which will be referred to as the `long exposure' $L'$ dataset, is used 
in \S 4 to investigate the nature of sources and blends 
outside of the central few arcsec of M31.

\subsection{CFHT Adaptive Optics Observations}

	Images of the central regions of M31 were recorded through $J, H,$ and $K'$ 
filters with the CFHT AO system and KIR imager on the night of September 10, 
2000. These data were used by Davidge (2001) to investigate the stellar content of the 
inner regions of the M31 bulge, and additional details of the instrument configuration, 
the observing procedures, and the reduction of these data can be found in that 
paper. To facilitate comparisons with the NIRI data, the CFHT images were 
re-sampled to have the same pixel scale as the NIRI $M'$ observations. 
A spatial filter was applied to simulate the removal of 
low spatial frequency bulge light that occured in the processing 
of the $M'$ and short exposure $L'$ data (see above), and the result was 
convolved with a gaussian to match the angular resolution of the $M'$ data.

\section{THE CENTRAL ARCSECOND OF M31 IN THE $1 - 5\mu$m WAVELENGTH INTERVAL}

	The morphology of the central arcsecond of M31 near 2.1 and $4.6\mu$m is 
investigated in Figure 1, where portions of the $K'$ (top panel) and $M'$ (middle 
panel) images are compared. The data have been rotated so that the line connecting P1 
and P2 falls along the horizontal axis. The nuclear morphologies 
at these wavelengths are very different. The brightest source in $K'$ is P1, and the 
elongation to the right of P1 due to P2$+$P3 is clearly visible. While P1 
also dominates in $M'$, the elongation due to P2$+$P3 is much more subtle than in $K'$.

	A more quantitative comparison of the wavelength dependence of the 
nuclear properties is conducted in Figure 2, where the light profiles in $J$, $K'$, $L'$ 
and $M'$ of a 0.6 arcsec wide strip along the P1 -- P2 axis are compared; the $L'$ 
light profile is that extracted from the short exposure $L'$ images. 
The light profiles in $J$ and $K'$ are almost indistinguishable 
after being scaled to the same peak P1 brightness; however, 
the $K'$ and $M'$ profiles are very different, in that 
the $M'$ profile falls below the $K'$ profile near P2. 
The $L'$ light profile also falls below the $K'$ profile near P2, although with a 
much smaller difference than is seen in the top panel.

	The differences between the $K'$ and $M'$ light profiles in Figure 2 are not due to 
a mis-match in image quality. Indeed, the image qualities of these data were matched 
using observations of standard stars that were recorded either before or after the 
datasets were obtained, and these were checked using a bright source that is 
7 arcsec south of the nucleus (\S 4). Comparisons between the $K'$ and $M'$ 
light profiles in which the image qualities of each were perturbed by 0.1 
arcsec with respect to the other indicate that the main result from the top panel of 
Figure 2, which is that the $K'$ and $M'$ light profiles are very different, is robust.

	The $M'$ light profile in Figure 2 can be modelled by combining the $K'$ 
profile with a point source that is slightly offset from the center of P1. 
This is demonstrated in the upper panel of Figure 3, where a scaled version of the 
$M'$ PSF, obtained from the standard star observations, has been added to the $K'$ 
profile. Initial efforts to combine the $K'$ profile and a 
scaled version of the PSF showed that the point source must be offset 
by $\sim 0.1$ arcsec from the center of P1, in the direction towards 
P2; with this offset, a reasonable match between the $M'$ and model profiles was 
found if the point source contributes $\sim 20\%$ of the peak light from P1 
in $M'$. The approximate integrated brightness of the point source is then $M' \sim 13$.

	The fit of this model to the $M'$ data perpendicular to the P1--P2 
axis provides an independent check of these results. The result of combining the 
$K'$ profile perpendicular to the P1 -- P2 axis with a point source that contributes 
20\% of the peak light from P1 is shown in the lower panel of Figure 3. 
It was not necessary to displace the point source from the centerline of 
the P1 -- P2 axis, indicating that the source likely falls within a few hundredths 
of an arc second of the P1 -- P2 axis. The model in which 20\% of the light comes 
from a point source provides a reasonably good match to the 
$M'$ profile perpendicular to the line connecting P1 and P2.

	A two dimensional representation of the model is shown 
in the lower panel of Figure 1. The simple model that we have considered provides 
a reasonable match to the observed morphology of the central arcsec 
of M31 in $M'$. Thus, we conclude that there is a source that is offset slightly from P1 
that is a major contributor to the light in the central arcsec of M31 near $4.6\mu$m.

	Various color profiles along the P1 -- P2 axis are shown in Figure 4. The $J-K$ and 
$K-L'$ colors are roughly constant in the vicinity of P1 and P2, with $J-K \sim 0.87$ 
and $K-L' \sim 0.2$, indicating that the integrated light in the $1 - 3\mu$m region 
of both components is dominated by K--M giants (e.g. Bessell \& Brett 
1988); this is consistent with the visible and near-infrared 
spectra of P1 and P2 (Kormendy \& Bender 1999). However, the $K-M'$ color varies along 
the axis connecting P1 and P2, in the sense that there is a bump that occurs 
between P1 and P2. The peak $K-M'$ color is 0.15 magnitudes redder than the region 
outside of this bump. The $K-M'$ color outside of the bump is roughly consistent with 
that expected for M giants, while the peak $K-M'$ color 
is redder than expected from a stellar photosphere. The nature of the source that 
contributes to the thermal emission near P1 is discussed further in \S 5.1.

	The CFHT data have FWHM $\sim 0.16$ arcsec in $H$ and $K'$ (Davidge 2001), 
and so have a markedly better angular resolution than the $L'$ and $M'$ data; for 
comparison, with FWHM $= 0.27$ arcsec, the angular resolution of the 
CFHT $J$ data is closer to that of the $L'$ and $M'$ data. 
The light and color profiles of the CFHT $H$ and $K'$ 
images in a 0.2 arcsecond wide strip along the P1 -- P2 axis,
without smoothing to match the angular resolution of the data at longer wavelengths, 
are shown in Figure 5. At this angular resolution P1 and P2 have comparable $H-K$ 
colors. There is also no signature of a red source near P1, 
indicating that the very red source of thermal emission near the 
center of P1 does not contribute significantly to the integrated near-infrared light.
Finally, there is no evidence in Figure 5 of a distinct component 
corresponding to P3. This is not surprising given the 
blue SED expected for such a young population and its modest size, both of which make 
P3 relatively faint in the near-infrared when compared with its 
surroundings. Indeed, the diminishing contribution that P3 makes to its surroundings 
towards longer wavelengths is clearly evident when comparing the F300W and 
F814W images shown by Lauer et al. (1998). 

\section{CIRCUMNUCLEAR SOURCES}

	The analysis in \S 3 was restricted to the central arcsecond of M31. However, 
the $L'$ and $M'$ observations also sample the inner bulge of M31 to distances in excess 
of 10 arcsec from the center of the galaxy. In this section we discuss the properties 
of sources that have been detected outside of the central arcsecond.

\subsection{A Bright AGB Star Detected in $M'$}

	One of the legacies of the IRAS Observatory was the discovery of 
sources with significant infrared emission in the direction of the inner Galaxy 
(Habing et al. 1985). The majority of these objects had not been 
detected at visible wavelengths at the time of their discovery with IRAS. The 
far-infrared flux was found to be variable in many of these 
sources, and Habing et al. (1985) concluded that they are likely 
evolved, dust-obscured stars. Subsequent observations found that the vast majority 
of these sources are Mira variables (e.g. Whitelock, Feast, \& Catchpole 1991; 
Allen, Kleinmann, \& Weinberg 1993). While the ages of these stars is a matter of 
controversy, as it depends on their assumed metallicity (e.g. Whitelock et al. 1991), 
they are potentially important probes of stellar 
content in heavily obscured regions, and/or very crowded 
environments, as they stand out against the background of less evolved objects 
when observed at thermal infrared wavelengths.

	The inner bulge of M31 may harbor a 
population of obscured AGB stars with photometric properties that 
are similar to the most evolved stars in the Galactic bulge. However, crowding 
is an obvious complicating factor when attempting to conduct a search for 
such objects; not only does crowding affect the faint limit of 
any photometric survey, but two sources of the same brightness that fall within 
the same angular resolution element will appear as a single source that is 0.75 mag 
brighter than either of the stars alone. While all stars detected in crowded 
environments like the central regions of M31 are blends, in many cases 
a bright star is blended with a much fainter object, and so the measured photometric 
properties are essentially those of the brighter star. Nevertheless, the incidence of 
blending between relatively bright objects is high enough that it has confounded 
efforts to study the stellar content of the M31 bulge using near-infrared data recorded 
during mediocre natural seeing conditions (e.g. Figure 15 of Stephens et al. 2003).

	The contrast between the relatively rare highly evolved bright red 
stars and the fainter, but much more numerous, blue objects increases 
as one moves to progressively longer wavelengths, and so 
the incidence of blends that have a marked impact on the photometric 
properties of the brightest red stars is reduced. A related observational 
consequence is that the amplitude of surface brightness flucuations increases with 
increasing wavelength, as predicted by models (e.g. Blakeslee, Vazdekis, 
\& Ajhar 2001; Liu, Charlot, \& Graham 2000; Jensen, Tonry, \& Luppino 1998; 
Worthey 1993). The contrast between bright stars 
that are embedded in dusty circumstellar envelopes and the underlying body of unresolved 
stars is enhanced even further at wavelengths longward of $2.5\mu$m, 
as stars in dust shells re-radiate much of their light at thermal infrared 
wavelengths, whereas the SEDs of stars without such envelopes are on the 
descending Rayleigh-Jeans side of the Planck distribution.

	Four bright candidate AGB stars can be identified in the circumnuclear 
regions of the $M'$ observations. The CFHT $K'$ data, which has the highest angular 
resolution of the datasets considered here, were used to determine if any of these 
objects could be associated with a single source, and thereby 
check that these objects are not the result of blends or noise spikes. 
Two of these objects are located in dense pockets of stars in the CFHT data, and so 
can not readily be identified with a single source. Another star did not have a 
corresponding bright source in the $K'$ image; this does not mean that this source 
is a noise event, as it could be a photometric variable that 
was at the lower end of its light variation when the $K'$ 
data were recorded. The remaining object could be identified 
with a single, isolated source, and the photometric properties of this object 
are listed in Table 1, along with offsets in right ascension and declination 
from the center of M31, in the sense that negative entries mean south and east 
of the nucleus; the $K-L'$ color uses the $L'$ brightness from the 
long exposure NIRI $L'$ image (\S 4.2).

	The object listed in Table 1 has $K-L'$ and $K-M' > 0$, and so is 
redder than M giants (e.g. Bessell \& Brett 1988); this suggests that the $M'$ light is 
a combination of photospheric and thermal components. An important caveat when assessing 
the $K-M'$ and $K-L'$ colors are that the $K$, $L'$ and $M'$ measurements were recorded 
during different epochs over a five year period. Very bright AGB stars can 
be photometrically variable, with amplitudes exceeding 1 
magnitude in the $2.5 - 5.0\mu$m wavelength interval, and so the colors may not be 
representative of the star at any given light curve phase. Nevertheless, the $H-K$ color 
listed in Table 1 is based on measurements that were separated from each other
by less than an hour, and is redder than what is expected for an M giant, for which 
$H-K$ is typically $0.2 - 0.3$. The $H-K$ color is thus also consistent with this 
star having significant amounts of circumstellar dust. 
The absolute $M'$ magnitude is $\sim -10$, which is consistent 
with the mean brightnesses of highly evolved AGB stars (e.g. Le Bertre 1992).
Thus, we conclude that the object listed in Table 1 is very likely a highly 
evolved AGB star. The properties of this star are discussed further in \S 5.2. 

\subsection{Circumnuclear Sources Detected in $L'$}

	Crowding is potentially more of an issue in $L'$ than it is in $M'$. 
There are a number of bright sources in the long exposure $L'$ 
image, and some of these are almost certainly blends of bright stars. 
A sample of objects that are not greatly affected by crowding was defined 
by examining the impact of degrading the angular resolution of the CFHT $K'$ image 
on measured brightness. This was done by convolving the $K'$ image 
with a gaussian so that it had the same angular resolution as the long exposure $L'$ 
image, and then using tasks in the DAOPHOT (Stetson 1987) and ALLSTAR (Stetson \& 
Harris 1988) packages to measure the brightnesses of sources in both the original and 
smoothed $K'$ images. The difference between the brightnesses in the smoothed and 
unsmoothed images, $\Delta K$, is a measure of the impact of blending; stars 
where blending has a significant impact on the photometry will have $\Delta K >> 0$, 
whereas sources that are not greatly affected by blending will have $\Delta K \sim 0$. 
It should be emphasized that this is a conservative method for identifying blended 
objects in $L'$, as the effects of blending are smaller in $L'$ than in $K'$.

	Given the high stellar density in this region of M31, it is not surprising that 
many of the objects are markedly brighter in the smoothed data than the unsmoothed 
data. Nevertheless, while $\Delta K$ is typically $\geq 0.3$ magnitudes, 
there are some stars with $\Delta K \sim 0$, and an inspection of the unsmoothed $K'$ 
images shows that the sources with $\Delta K < 0.15$ tend to be isolated, 
in the sense of not having companions of comparable brightness within a few tenths of 
an arcsecond, and have round image cores. Therefore, sources with $\Delta K < 0.15$ 
magnitudes are identified as not being severly affected by crowding. 

	The brightnesses of stars in the long exposure $L'$ image were measured using 
tasks in DAOPHOT and ALLSTAR, and the $(L', K-L')$ CMD of sources with $\Delta K < 0.15$ 
is shown in Figure 6. The majority of stars have $M_{L'} > -9.6$ and $K-L'$ 
between 0 and 2. The object with M$_{L'} \sim -10.5$ is the AGB star discussed in \S 4.1; 
this object has $\Delta K = 0.05$. Points corresponding to luminous M giants in the 
Galactic bulge, based on the photometry presented in Table 1 of 
Frogel \& Whitford (1987), as well as the Galactic disk C stars studied 
by Le Bertre (1992), are shown in the right hand panel of Figure 6. \footnote[3]
{Frogel \& Whitford (1987) used the $L$ filter for their observations. A comparison of 
the $L$ and $L'$ brightnesses of UKIRT bright standard stars 
(http:/www.jach.hawaii.edu/UKIRT/astronomy/calib/phot\_cal/ukirt\_stds.html) shows that 
on average $<L - L'> = -0.01 \pm 0.01$ for K and M giants/supergiants, 
where the quoted uncertainty is the standard error in the mean. Given this small 
offset, the Frogel \& Whitford $L$ measurements are assumed to be in the $L'$ system.}

	There are three main conclusions that can be reached from Figure 6. First, the 
majority of unblended stars in our sample have intrinsic brightnesses that agree to within 
$\sim 0.3$ magnitudes with the brightest M giants in the Galactic bulge. This 
difference is comparable to the uncertainties introduced by depth effects in the Galactic 
bulge and errors in the photometric calibration and residual blending. Second, the 
AGB star discussed in \S 4.1 is clearly brighter than the majority of 
stars in the Galactic Bulge. Third, the majority of C stars in the Le Bertre (1992) 
sample have $K-L' > 2$, and hence are redder than the brightest stars in the Galactic 
bulge and the inner regions of the M31 bulge. The comparison in Figure 6 
confirms the notion that the brightest AGB stars in the bulge of M31 and the 
Milky-Way are similar, both in terms of peak brightness and color (e.g. Davidge 2001).

	IRS 7 is the most luminous source in the SgrA complex, and so is an interesting 
benchmark for stellar content studies of the central regions of other galaxies. 
The location of IRS 7 on the $(M_{L'}, L'-K)$ CMD is shown in Figure 6. 
The points plotted for IRS 7 are based on photometry given by Blum, 
Sellgren, \& DePoy (1996) and Viehmann et al. (2005). Whereas these two studies find 
similar $L'$ magnitudes, they give very different $K-L'$ colors, perhaps due to photometric 
variability (e.g. Blum et al. 1996). 

	A source as bright as IRS 7 would clearly be seen in our $L'$ data 
if it were in an uncrowded environment, and no counterpart is detected. This does not 
mean that a source as bright as IRS 7 is not present, as sources of this brightness 
that happen to be close to other bright stars may be rejected as blends in the 
smoothing analysis used here to identify isolated stars. The source near P1 
discussed in \S 3 likely has a luminosity that is comparable to that of IRS 7, although it 
is not detected in $L'$ because it is faint at this wavelength (e.g. Figure 2), and is 
located in an extremely crowded environment.

\section{DISCUSSION \& SUMMARY}

	Images recorded through $L'$ and $M'$ filters with NIRI on Gemini North have 
been used to investigate the central regions of M31. The data have been combined with 
the high-angular resolution near-infrared images recorded with the CFHT AO system that 
were discussed previously by Davidge (2001). The $3 - 5\mu$m wavelength region is 
important for stellar content studies since stars with dusty circumstellar 
envelopes can be identified at these wavelengths in relatively crowded environments. 
Despite the obvious problems caused by high background levels that plague 
ground-based observations in the thermal infrared, with large ground-based telescopes 
it is possible to obtain data with an angular resolution 
that exceeds what can be achieved with smaller aperture facilities in space.

\subsection{The Nuclear Regions of M31}

	As one moves from visible to infrared wavelengths, the 
dominant source of light from composite stellar systems shifts from 
stars near the main sequence turn-off to progressively more evolved objects. 
One consequence is that age-related differences in stellar content 
can be more difficult to detect in the infrared than at visible wavelengths, as a much 
longer period of time is required to produce measurable changes in the properties 
of -- for example -- RGB stars than main sequence stars. This is evident 
in the central regions of M31, where the near-infrared colors of P1 and P2 are similar 
(Figures 4 and 5), even though there are obvious differences between their SEDs at shorter 
wavelengths (Lauer et al. 1998), due to the moderately young nuclear star cluster P3.

	At wavelengths longward of $2.5\mu$m the SEDs of composite systems are 
very insensitive to star formation history, unless there is extremely active 
star formation (e.g. Hunt, Giovanardi, \& Helou 2002); in fact, the SEDs of galaxies 
covering a diverse mix of morphologies and star-forming 
histories converge at wavelengths longward of $3\mu$m (e.g. Figure 3a of Boselli 
et al. 2003). Given this tendency for SEDs to converge at thermal wavelengths, any 
large-scale differences in the SEDs of stars and stellar systems in the $3 
- 5\mu$m region will likely be due to the presence of non-photospheric light. 
Indeed, the SEDs of Virgo cluster galaxies shown in Figure 3 of Boselli et al. (2003) 
indicate that galaxies with very different morphological types (and stellar contents) have 
$K - M' \sim -0.1$ with a type-to-type scatter of only a few 
hundredths of a magnitude. This is similar to the $K-M'$ color that we measure throughout 
most of P1 and P2. Given the modest scatter in $K-M'$ colors, a source that is 0.15 mag 
redder in $K-M'$ than the bulk of the population can not be obtained simply by 
combining the spectra of `normal' composite stellar systems; rather, a 
non-photospheric thermal component must be present.

	In the case of M31, P1 and P2 might be expected to 
have different SEDs at wavelengths longward of $2.5\mu$m if there is 
emission with characteristic temperatures $100 - 1000$ K from sources in P3. 
However, even though thermal emission from dust is detected from many sources near Sgr A 
(Viehmann et al. 2005), the amount of such emission from P3 is expected to be modest. Bender 
et al. (2005) find that P3 has an age $\sim 0.2$ Gyr and a mass 5200 M$_{\odot}$. While 
individual stars have not yet been resolved in P3, the simulated CMD shown in their 
Figure 6 consists mainly of main sequence stars later than B5, with only 7 evolved 
red giants. Thus, P3 likely lacks the massive main sequence stars or extremely 
bright AGB stars that are sites of thermal emission near the Galactic Center.
Our results are consistent with this expectation, as we 
do not detect a red bump in the $H-K$ and $J-K$ color profile near P2,  
suggesting that P3 does not contain a large grouping of bright cool sources. Moreover, 
in \S 3 it is shown that the $K-M'$ color of P2 $+$ P3 is consistent with 
photospheric light with no thermal emission.

	What is the nature of the $M'$-bright source 
near P1? In \S 3 it is demonstrated that the morphological 
properties of the central arcsec of M31 in $M'$ can be modelled by 
combining the $2.1\mu$m light profile, where the peak brightnesses of 
P1 and P2 agree to within a few tenths of a magnitude, with a single bright source, that 
is offset slightly from the center of P1. The source contributes $\sim 20\%$ of the 
light from P1, and has a brightness M$_{M'} \sim -11.5$. That this object has only a 
very minor impact on the $L'$ light profile in Figure 2 suggests that the emission 
can be characterised by an effective temperature of only a few hundred K. The emission 
occurs 0.35 arcsec away from P2, which corresponds to a projected distance of $\sim 
1.3$ parsecs, and so is likely not associated with the central black hole.

	The disk model for P1 and P2 forwarded by 
Tremaine (1995) has been highly successful in reproducing the dynamical 
properties of this system (e.g. Bender et al. 2005; Kormendy \& Bender 1999), 
and predicts that P1 and P2 should have similar stellar contents. 
The results discussed here do not pose a problem for the Tremaine 
(1995) model, as the emission detected in the $M'$ data may come from a single 
star that is extremely bright in $M'$. Indeed, 
the bulge of M31 was detected by IRAS, and the 12 and $25\mu$m fluxes indicate 
that circumstellar emission from warm dust is present (Soifer et al. 1986); consequently, 
individual sources with red $K-M'$ colors might be expected in 
suitably deep $M'$ images. The intrinsic $M'$ brightness 
estimated from the models in \S 3 is consistent with what is seen 
among heavily obscured, intrinsically luminous AGB stars near the peak of their 
light curves. Moreover, while the majority of circumstellar shells around AGB stars have 
effective temperatures close to 1000 K some, such as AFGC 3068, have dust shells that 
are as cool as 300 K (e.g. Le Bertre 1997). 

	Do we expect to find a single star of this brightness in the central arcsec of M31? 
To answer this question, we compute R$_{1AGB}$, which is the radius 
that will contain one star in the range of progenitor masses for AGB stars with a brightness 
like the source near P1. To compute R$_{1AGB}$ it is necessary to 
make assumptions about the stellar content near the center of M31 and the properties of 
the AGB population. We assume for the baseline calculation (1) that the stars near the 
center of M31 have roughly solar metallicities and (2) that the star-forming history 
near the center of M31 averaged over the past 1 Gyr, which is the approximate maximum age of 
AGB stars of the brightness we are considering, has been similar to that near 
the Galactic Center, where Blum et al. (2003) find that $\sim 1\%$ of the stellar 
mass has formed within the past 1 Gyr. We further assume that AGB stars 
near the center of M31 have $K-M'$ colors that are similar to the most highly 
evolved stars near the Galactic Center. Blum et al. (2003) obtained spectra of a number of 
bright sources near the Galactic Center, and the brightest spectroscopically confirmed 
AGB stars in their sample that have $M'$ photometric measurements in Viehmann et al. 
(2005) are IRS 9 and IRS 12N. If A$_K = 3.1$ magnitudes (Davidge 1998), then 
IRS 9 and IRS 12N have $(K - M')_0 \sim 1.0 - 1.3$. The location of these stars 
on the Herzsprung-Russell diagram suggest that they likely 
have an age $\sim 1$ Gyr (Blum et al. 2003). Given that the source near P1 is 
likely younger than this (see below), and that younger AGB stars will be more 
massive and so may have redder $K-M'$ colors, we adopt the upper value $(K-M')_0 = 
1.3$ for our baseline calculation.

	Given that highly evolved AGB stars may be LPVs with amplitudes $\sim 1$ magnitude 
in the infrared, then the source near P1 may have a mean M$_{M'}$ as low as $\sim -10.5$, 
which corresponds to M$_K = -9.2$ if $K-M' = 1.3$; thus, we examine the 
statistics of AGB stars with M$_K \leq -9.2$ for the baseline model. Moreover, we 
consider a population with log(t$_{yr}$) = 8.1 -- this is the age at which the 
AGB phase transition occurs, so that the most luminous AGB stars will be present. 
While adopting such a favourable age may seem contrived, there is evidence for stars close 
to this age in the central arcsec of M31 (Bender et 
al. 2005). With the age and metallicity thus fixed, the 
isochrones tabulated by Girardi et al. (2002) can be used to find the progenitor masses of 
AGB stars with M$_K \leq -9.2$. The fractional contribution that these stars make 
to the total system mass for a given mass function can then be computed.

	We find that a solar metallicity system with an age log(t$_{yr}$) = 8.1 and 
a Miller \& Scalo (1979) mass function must have a mass of $3 \times 10^4$ M$_{\odot}$ to 
have one star in the required mass range. Adopting (M/L)$_K = 0.2$ for a population with 
this age (Mouhcine \& Lancon 2002), then such a system will have M$_K = -10$. If this 
population accounts for 1\% of the total mass near the center of M31 then the 
$K-$band surface brightness profile constructed from NIRI data (Jensen 2006, in 
preparation) indicates that R$_{1AGB} = 1$ arcsec; thus, we might expect 
to find a suitably bright AGB star within the central arcsec of M31.

	To be sure, the calculation of R$_{1AGB}$ is fraught with uncertainties. 
For example, if the intermediate-age component makes up 0.1\% of the total mass, rather 
than 1\%, then R$_{1AGB} \sim 6$ arcsec, although there would still be a 10\% chance of 
detecting a suitably bright AGB star in the central arcsec. If an age log(t$_{yr}$) = 8.2 is 
adopted then R$_{1AGB}$ grows by 0.5 arcsec. If a redder $K-M'$ color is assumed, 
which is not unreasonable if the stars are younger than IRS 9 and IRS 12N, then 
R$_{1AGB}$ is also altered and a larger range of ages can be considered. For example, 
if $K-M' = 2$ (roughly midway between the $K-M'$ colors of 
IRS 9 $+$ 12N and IRS 7) then R$_{1AGB}$ that is computed assuming log(t$_{yr}$) = 8.2 drops 
to 1 arcsec. If a Salpeter (1955) mass function is adopted then 
larger R$_{1AGB}$ values are found, with R$_{1AGB} \sim 5.5$ arcsec for the baseline 
assumptions. Finally, there are also uncertainties 
in the isochrones themselves, which we have not considered. 
Given the various uncertainties, we feel that the most robust conclusion that 
can be drawn is that the discovery of a single bright AGB star in the central arcsec of 
M31 is not unexpected given the wide range of plausible assumptions about the 
progenitor stars and the contribution that they may make to the total stellar mass.

	The model in which a single bright AGB star is responsible 
for producing the emission near P1 can be tested in three 
ways. First, low to moderate resolution spectra in the $5\mu$m atmospheric window 
with sub-arcsecond angular resolution could be used to search for continuum 
emission from hot dust. The CO fundamental bands, which fall 
between 4.6 and $4.7\mu$m, are prominent absorption features 
in the spectra of late-type stars. Continuum emission from hot dust will veil these 
features, and so we predict a variation in the integrated CO band strengths along the 
line connecting P1 and P2. The relative $K-M'$ colors of P1 and P2 predict that the 
fundamental CO bands will be weakest 0.1 arcsec away from P1. While the first overtone CO 
bands near $2.3\mu$m occur at wavelengths where the sky background is lower, and hence 
would be easier to observe, the effective temperature of the thermal emission is such that 
a detectable thermal continuum is not expected at near-infrared wavelengths.

	A second test is to search for changes in the brightness of the source near P1. 
Many bright AGB stars are photometric variables, with 
periods on the order of hundreds of days and amplitudes 
approaching $\sim 1$ magnitude in $M'$. Therefore, if the single star interpretation is 
correct, then the $M'$ flux may change with time, with measureable differences 
expected from year to year. This raises the intriguing possibility that the 
$M'$ morphology of the M31 nucleus may change over moderately short time scales. 

	A third test is to check that the light comes from a point source, 
rather than from an extended source, such as a dust-enshrouded star cluster. Clusters
near the center of the Milky-Way have sizes $\sim 1$ parsec, and this corresponds to 
an angular size of $\sim 0.3$ arcsec when viewed at the distance of M31.
For comparison, the $M'$ data used here have FWHM $\sim 0.4$ arcsec. If 
images recorded with higher angular resolution find that the source is not 
extended, then it will place tighter constraints on it being a single object. 

\subsection{Circumnuclear Sources}

	Using the CFHT $K$ images as a guide, a population of isolated stars has 
been identified in $L'$. The brightest AGB stars in the bulges of the Milky-Way and M31 
have similar peak $K-$band brightnesses (Davidge 2001; Davidge et al. 2005). The data 
discussed in the present paper further strengthens the notion that the dominant AGB 
populations in the bulges of the two galaxies are similar, in that the peak $L'$ 
brightnesses of objects in the bulges of the Milky-Way and M31 agree to $\sim 0.3$ 
magnitudes, which is within the uncertainties introduced by the photometric calibration 
and depth effects in the Galactic bulge.

	While a number of sources have been identified in our $L'$ data, only a single 
source, which has relatively red $K-L'$ and $K-M'$ colors, has been detected 
outside of the nuclear regions of M31 in $M'$. The colors listed in Table 1 were computed 
from data recorded at different epochs, and bright AGB stars can be variable with 
amplitudes approaching one magnitude in the infrared. This could explain why the 
$K-L'$ color in Table 1 is larger than the $K-M'$ color, although there are stars 
such as IRC --30060 that have $K-L' > K-M'$ (Le Bertre 1992). Nevertheless, 
near simultaneous $H$ and $K$ measurements give an 
$H-K$ color that is consistent with a reddened source. 
The object is a round point source on the CFHT $K$ image, which has 
an angular resolution of 0.15 arcsec ($\sim 0.6$ parsecs), and so it 
is unlikely that it is a cluster of very young stars in a common dust envelope. Rather, 
it is more likely that -- like the source found near P1 -- it is a star 
evolving near the peak of the AGB and has a circumstellar envelope. Indeed, 
with M$_{M'} \sim -10$ this star has a $5\mu$m brightness that is typical of the AGB 
stars observed by Le Bertre (1992). This object is $\sim 1.5$ magnitude fainter in 
$M'$ than the source detected near P1; nevertheless, a direct comparison of $M'$ 
brightness is complicated by photometric variability and possible differences in the 
optical depth of the circumstellar envelopes. 

	The star discussed in the previous paragraph has M$_K < -9$, and hence is close to 
the peak brightness expected for AGB stars. In fact, isochrones from Girardi et al. (2002) 
predict that the brightest AGB stars with solar metallicity have M$_K \sim -9.9$ when the 
age is 0.13 Gyr, whereas for systems with ages of 1 and 10 Gyr the peak AGB brightnesses 
are M$_K = -8.6$ and --7.5; the predicted peak brightnesses change little if a 
more metal-rich composition is adopted. Keeping in mind the caveat of stellar 
variablity, the star detected in $M'$ may thus belong to a population with 
an age of a few 100 Myr. This source has a projected distance of 30 parsecs 
from the center of M31, and it is tempting to associate it with the most 
recent episode of nuclear star formation, which Bender et al. (2005) suggest 
occured 200 Myr in the past. 

	The presence of a luminous intermediate age star with a projected distance 
of $\sim 30$ parsecs from the center of M31 might suggest that wide scale star formation 
occured throughout the inner bulge of M31 a few hundred Myr in the past. However, 
the star may have formed closer to the nucleus of M31. Indeed, it has 
been suggested that dynamical interactions with giant molecular clouds and the tidal 
disruption of young clusters may scatter stars into 
the Galactic bulge (e.g. Figer et al. 1999; Kim \& Morris 2001), and similar 
processes could be at work near the center of M31. Details of the birthplace of this 
source notwithstanding, if it belongs to a centrally-concentrated population then 
stars with a similar luminosity will not be uniformly mixed throughout the M31 bulge, 
and a larger field survey of the inner regions of M31 in $M'$ would find that the 
distribution of objects with similar luminosity does not follow the 
surface brightness profile of the bulge.

	Could this object be a planetary nebulae (PN)? Planetary nebula can have very red 
$K-M$ colors (e.g. Phillips \& Cuesta 1994). However, the object in 
Table 1 is likely too bright to be a PN. Indeed, the PNe seen towards the Galactic bulge 
tend to have $K$ between 11.5 and 13.0 (M$_K$ between --1.5 and --3.0), although PNe as 
bright as M$_K \sim -7.5$ may also be present (e.g. Ramos Larios \& Phillips 2005). 
For comparison, the object in Table 1 has M$_K \leq -9$. Finally, PNe have 
$K-M >> 1$ (Phillips \& Cuesta 1994), which is not the 
case for the object detected here. Hence, it is likely not a PN.

\acknowledgements{It is a pleasure to thank the anonymous referee for providing thoughtful 
comments that helped improve the paper.}

\clearpage

\parindent=0.0cm

\begin{table*}
\begin{center}
\begin{tabular}{ccccccc}
\tableline\tableline
$\Delta \alpha$ & $\Delta \delta$ & $K$ & $M'$ & $K-M'$ & $K-L'$ & $H-K$ \\
 (arcsec) & (arcsec) & & & & & \\
\hline
--1 & --8 & 14.5 & 14.3 & 0.2 & 0.5 & 0.8 \\
\tableline
\end{tabular}
\end{center}
\caption{Source Detected in $M'$}
\end{table*}

\clearpage

\begin{figure}
\figurenum{1}
\epsscale{0.5}
\plotone{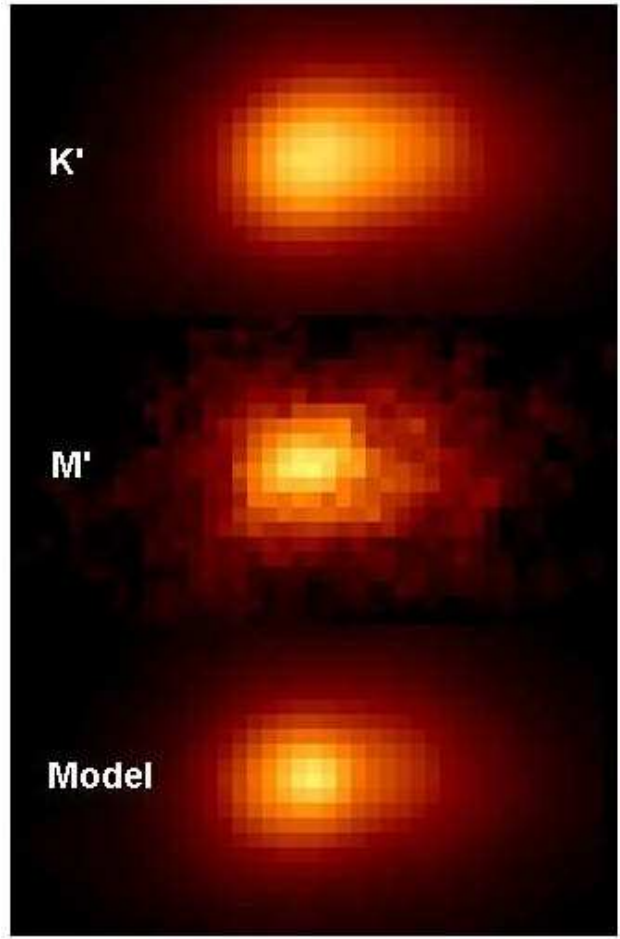}
\caption
{The $K'$ (top) and $M'$ (middle) images of the center of M31. A $2 \times 4$ arcsec 
region is displayed for each filter, and the data have been rotated so that the line 
connecting P1 and P2 falls along the horizontal axis. The images are displayed with the 
same stretch after being normalized to the same peak P1 brightness. Note that while P1 
and P2 form an elongated structure in $K'$, P1 is more dominant with respect to P2 in $M'$. 
The bottom image is the model described in \S 3, in which a point source, offset by 
0.1 arcsec from the center of P1 along the P1 -- P2 axis and contributing 20\% of 
the peak light from P1, has been added to the $K'$ image. This simple model reproduces 
the overall appearance of the center of M31 near $4.6\mu$m.}
\end{figure}

\begin{figure}
\figurenum{2}
\epsscale{1.0}
\plotone{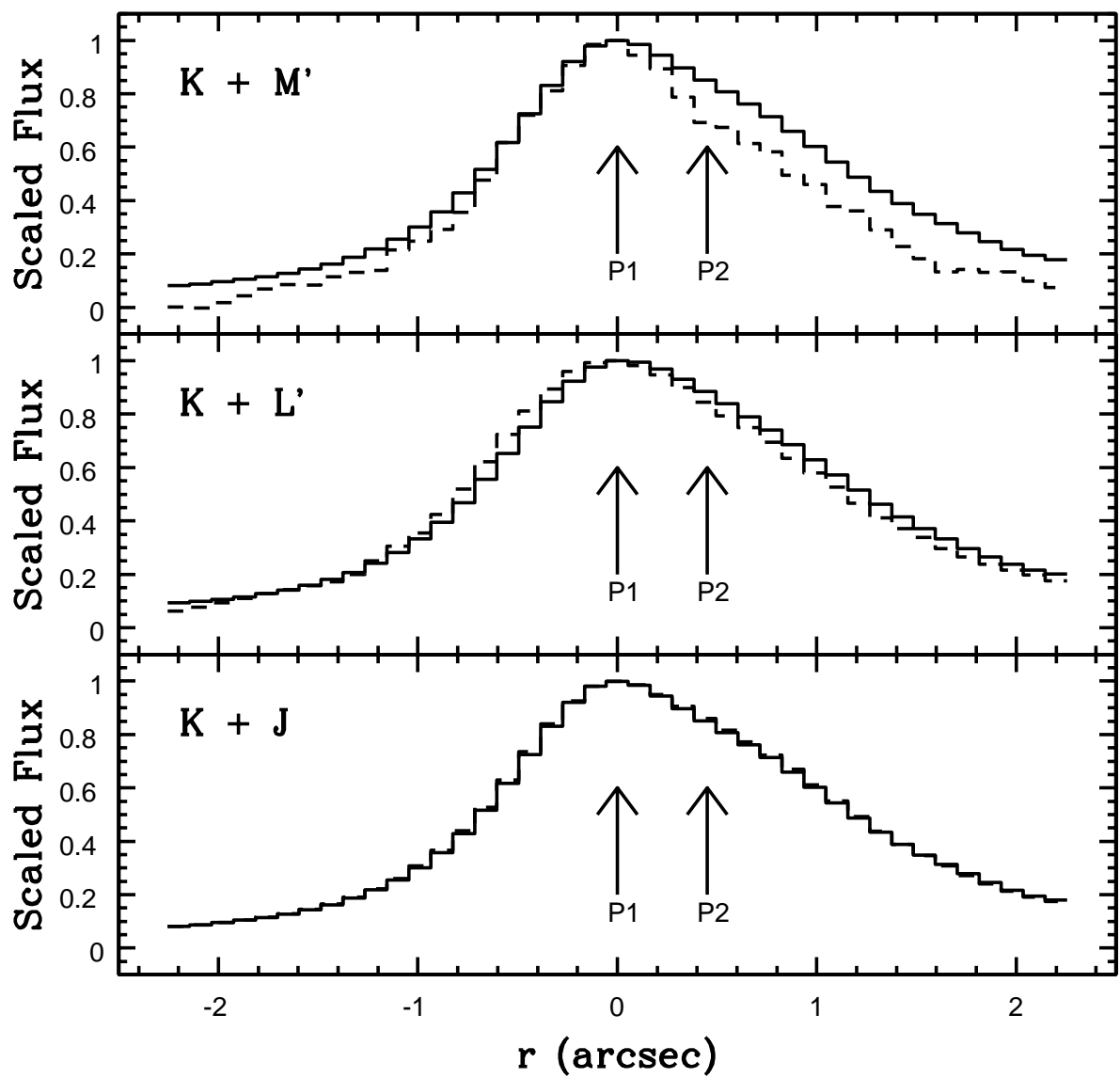}
\caption
{Light profiles along the P1 -- P2 axis. The curves show the average flux in a 0.6 arcsec 
wide strip; $r$ is the distance in arcsec from the center of P1. The $K'$ light 
profile is the solid line in each panel, while the light profiles in $J$ (lower 
panel), $L'$ (middle panel), and $M'$ (upper panel) are 
shown as dashed lines. All profiles have been normalized according to the peak P1 
value, and the $K'$ profile in the middle panel has been smoothed slightly 
to match the angular resolution of the $L'$ data. The $M'$ profile falls below 
the $K'$ profile when $r > 0.3$ arcsec. This same behaviour is also seen, 
although to a lesser extent, in $L'$.} 
\end{figure}

\begin{figure}
\figurenum{3}
\epsscale{1.0}
\plotone{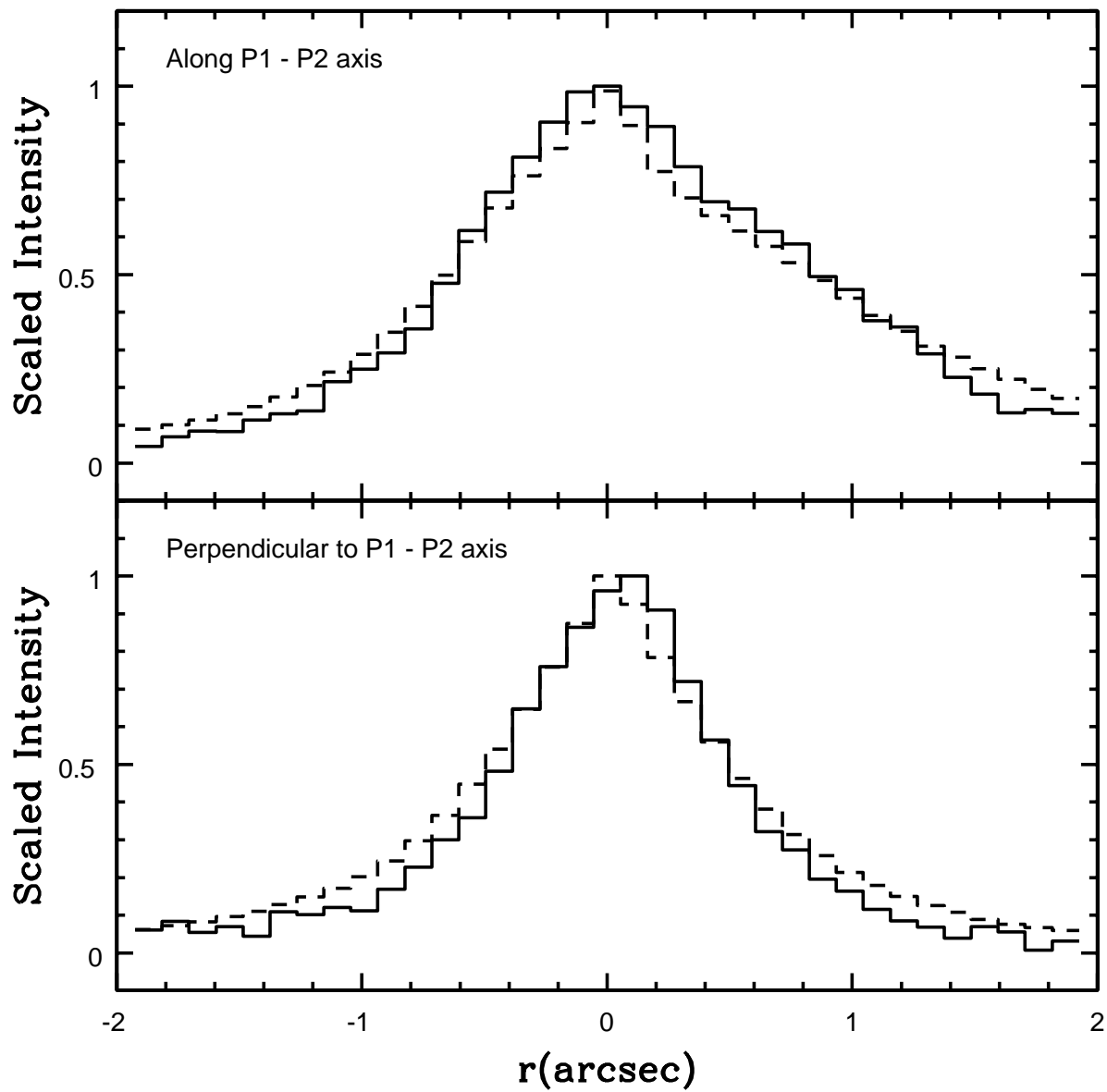}
\caption
{The results of adding a point source to the $K'$ light profiles along the 
P1--P2 axis (top panel), and perpendicular to the P1--P2 axis (lower panel). 
The solid line is the $M'$ light profile, while the dashed line in each panel 
shows the $K'$ light profile combined with a source that contributes 
20\% of the peak flux; in the upper panel the source has been shifted by 0.1 arcsec 
along the P1 -- P2 axis towards P2. Note that the addition of a point source to the 
$K'$ light profile can account for the differences seen in the upper panel of Figure 2.}
\end{figure}

\begin{figure}
\figurenum{4}
\epsscale{1.0}
\plotone{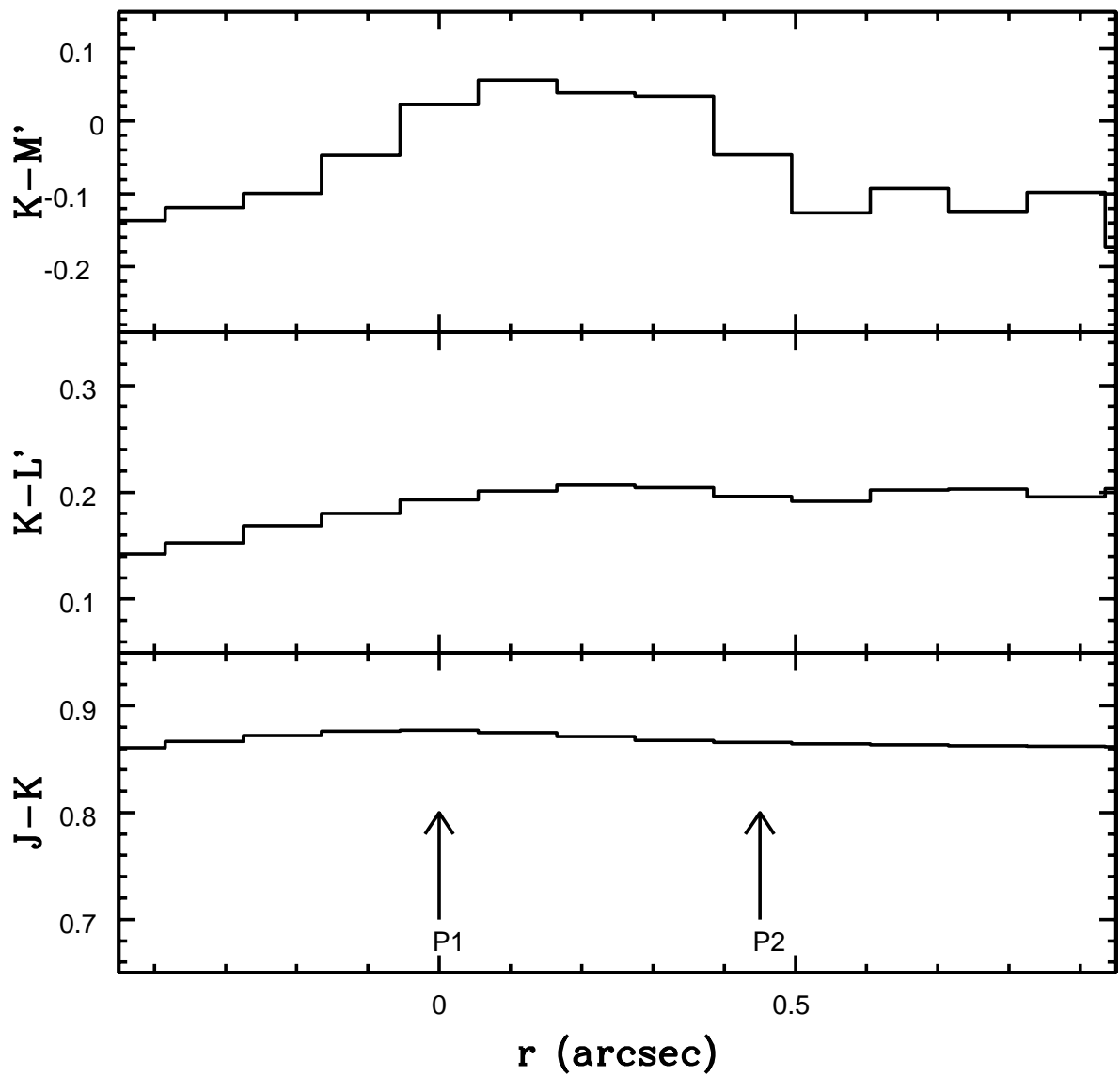}
\caption
{Color profiles along the P1--P2 axis, constructed from 
the profiles shown in Figure 2. The $K'$ data used to construct the $K-L'$ profile 
have been smoothed to match the slightly lower angular resolution of the $L'$ data with 
respect to the $M'$ data. Note that whereas P1 and P2 have comparable $J-K$ and 
$K-L'$ colors, there is a bump in the $K-M'$ profile between P1 and P2. 
This suggests that there is a source of thermal emission $\sim 0.1$ arcsec from 
the center of P1 that has an effective temperature of a few hundred K.}
\end{figure}

\begin{figure}
\figurenum{5}
\epsscale{1.0}
\plotone{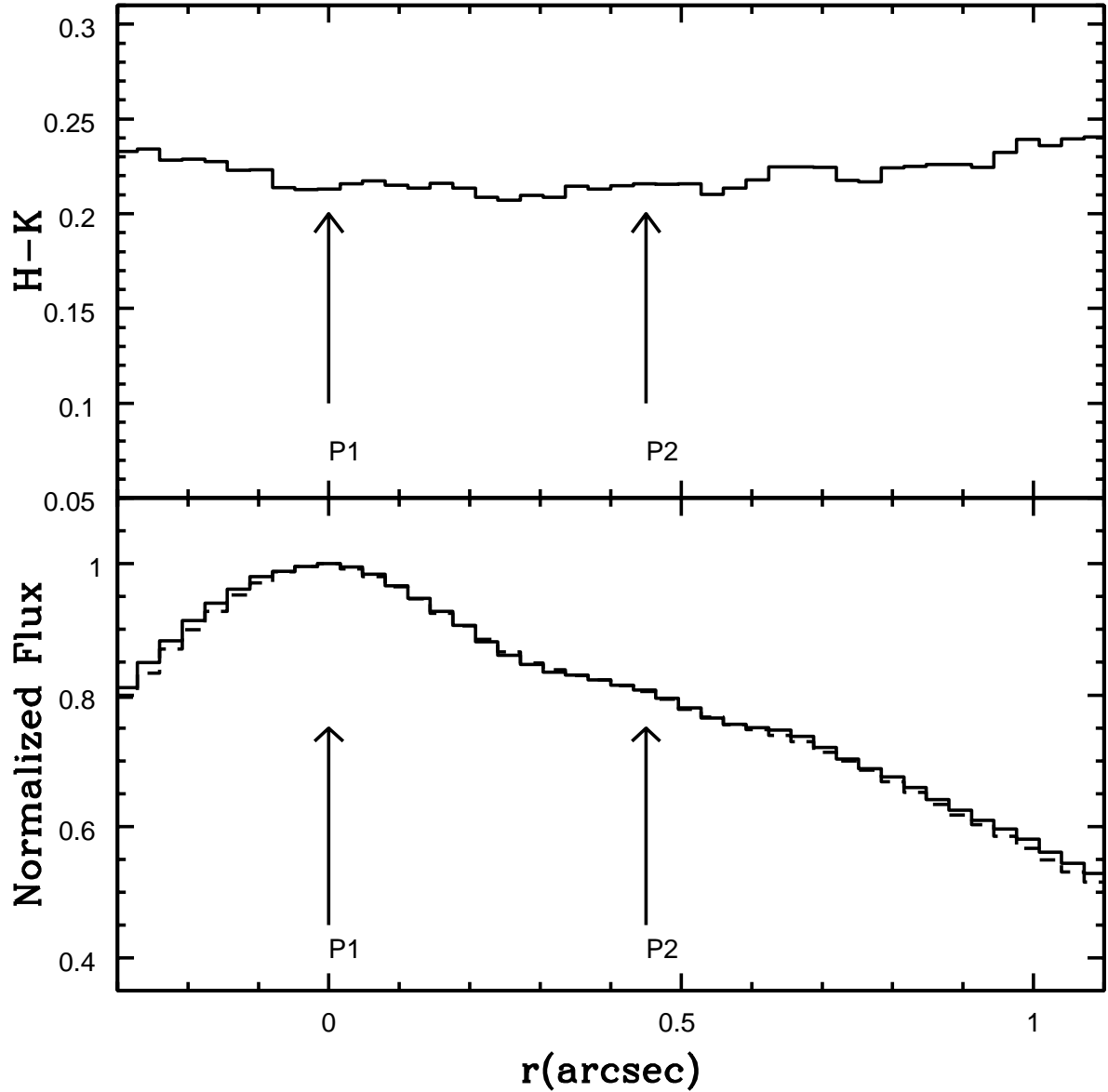}
\caption
{Light and color profiles of the CFHT $H$ and $K'$ data along the P1 -- P2 axis, both of 
which have an angular resolution of 0.16 arcsec FWHM. The profiles show 
the average flux in a 0.2 arcsec wide strip. Both the $K'$ 
(solid line) and $H$ (dashed line) profiles in the lower panel have been 
normalized to the peak signal at P1. Note that P1 and P2 have very similar $H-K$ 
colors.}
\end{figure}

\begin{figure}
\figurenum{6}
\epsscale{0.8}
\plotone{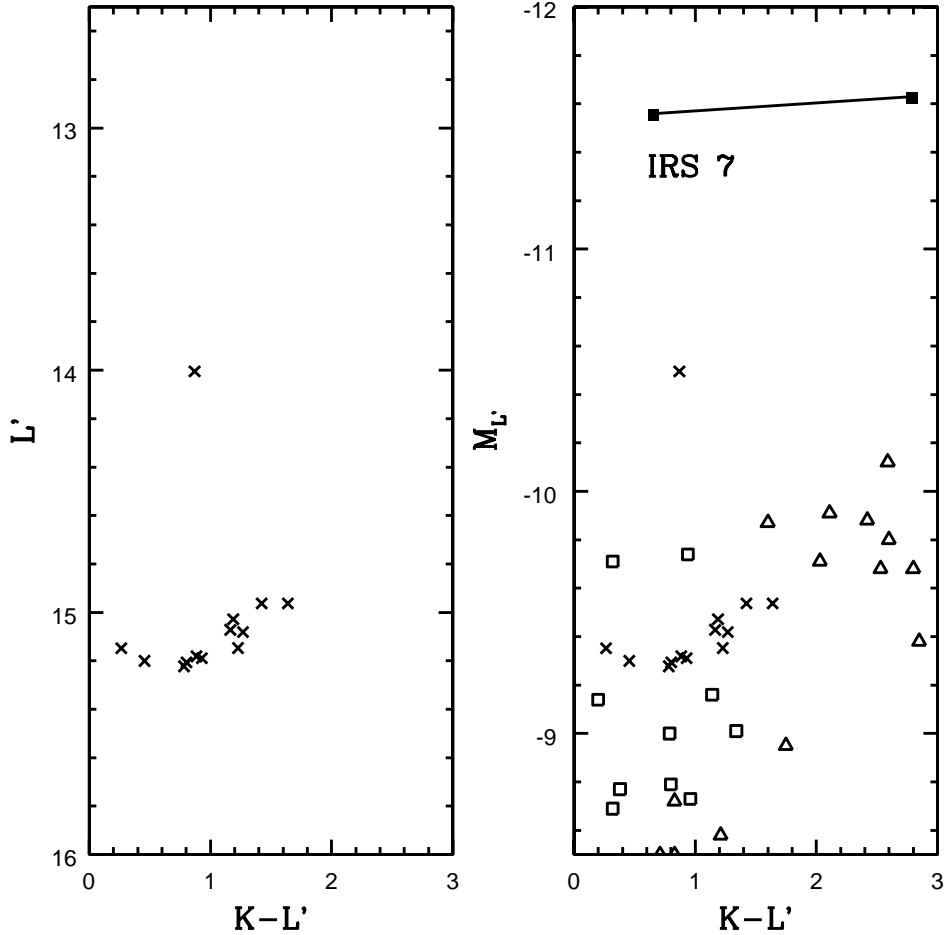}
\caption
{The $(L', K-L')$ and $(M_{L'}, K-L')$ CMDs of the central regions of M31. Stars in M31, 
which were found not to be significantly affected by blending based on the procedure 
discussed in \S 4, are shown as crosses. A distance modulus of 24.47 (Stanek 
\& Garnavich 1998), which is based on the brightness of the red horizontal branch (HB) 
clump, has been assumed for M31. The filled squares show the position 
of IRS 7 based on photometry from Blum, Sellgren, \& DePoy (1996) and Viehmann 
et al. (2005); while these two studies find very similar $L$ magnitudes, 
they measure very different $K-L$ colors. For consistency with the M31 
distance, a distance modulus of 14.57 (Stanke \& Garnavich 1998), 
which is also based on the HB clump, has been assumed for the Galactic 
Center. The reddening towards IRS 7 was assumed to be A$_K = 3.1$ 
(Davidge 1998), and the reddening curve of Rieke \& Lebofsky (1985) has been adopted.
The open squares show M giants in Baade's Window from Table 1 of Frogel \& 
Whitford (1987), while the open triangles show the mean brightnesses and colors 
of the C stars studied by Le Bertre (1992); the adopted distances for these stars are 
those listed in Table 1 of Le Bertre (1997). Note that the $K-L'$ colors of the majority 
of objects detected in M31 match those seen among M giants in the Galactic bulge, and that 
the peak M$_{L'}$ brightnesses of the brightest stars in the M31 and Galactic 
bulges differ by only $\sim 0.3$ magnitudes when $K-L' < 2$.}
\end{figure}

\end{document}